\documentclass[a4paper,conference]{IEEEtran}
\usepackage{times,amsmath,amssymb}
\usepackage{slashbox}
\usepackage{graphicx}
\usepackage{xcolor}
\usepackage{algorithm2e}
\usepackage{algorithmicx}
\usepackage{setspace}
\usepackage[export]{adjustbox}
\usepackage{multirow}
\usepackage{bm}
\usepackage[noend]{algpseudocode}
\usepackage{stmaryrd}

\begin{document}

\title{Data Normalization for Bilinear Structures in High-Frequency Financial Time-series}
\author{\IEEEauthorblockN{Dat Thanh Tran\IEEEauthorrefmark{1}, Juho Kanniainen\IEEEauthorrefmark{1}, Moncef Gabbouj\IEEEauthorrefmark{1}, Alexandros Iosifidis\IEEEauthorrefmark{2}}
\IEEEauthorblockA{\IEEEauthorrefmark{1}Department of Computing Sciences, Tampere University, Finland\\
\IEEEauthorrefmark{2}Department of Engineering, Aarhus University, Denmark\\
Email:\{thanh.tran, juho.kanniainen, moncef.gabbouj\}@tuni.fi, alexandros.iosifidis@eng.au.dk}\\

}

\maketitle

\begin{abstract}
Financial time-series analysis and forecasting have been extensively studied over the past decades, yet still remain as a very challenging research topic. Since the financial market is inherently noisy and stochastic, a majority of financial time-series of interests are non-stationary, and often obtained from different modalities. This property presents great challenges and can significantly affect the performance of the subsequent analysis/forecasting steps. Recently, the Temporal Attention augmented Bilinear Layer (TABL) has shown great performances in tackling financial forecasting problems. In this paper, by taking into account the nature of bilinear projections in TABL networks, we propose Bilinear Normalization (BiN), a simple, yet efficient normalization layer to be incorporated into TABL networks to tackle potential problems posed by non-stationarity and multimodalities in the input series. Our experiments using a large scale Limit Order Book (LOB) consisting of more than 4 million order events show that BiN-TABL outperforms TABL networks using other state-of-the-arts normalization schemes by a large margin. 
\end{abstract}

\section{Introduction}\label{S:Intro}
Although we have observed great successes in time-series and sequence analysis \cite{tran2018temporal, zhang2019deeplob, passalis2018temporal, tran2017multilinear, tran2017tensor, passalis2017time}, and the topic in general has been extensively studied, we still face great challenges when working with multivariate time-series obtained from financial markets, especially high-frequency data. In High-Frequency Trading (HFT), traders focus on short-term investment horizon and profit from small margin of the price changes with large volume. Thus, HFT traders rely on market volatility to make profit. This, however, also poses great challenges when dealing with the data obtained in the HFT market. 

Due to the unique characteristics of financial market, we still need a great amount of efforts in order to have the same successes as in Computer Vision (CV) \cite{ren2015faster, redmon2016you,iosifidis2012view, tran2018improving, tran2019multilinear, tran2019bmultilinear} and Natural Language Processing (NLP) \cite{bahdanau2014neural, devlin2018bert}. On one hand, the problems targeted in CV and NLP mainly involve cognitive tasks, whose inputs are intuitive and innate for human being to visualize and interpret such as images or languages while it is not our natural ability to interpret financial time-series. On the other hand, images or audio are well-behaved signals in a sense that the range or variances are known or can be easily manipulated without loosing the characteristic of the signal, while financial observations can exhibit drastic changes over time or even at the same time instance, signals from different modalities can be very different such as the stock prices. Thus data preprocessing plays an important role when working with financial data. 

Perhaps the most popular normalization scheme for time-series is z-score normalization, i.e. transforming the data to have zero-mean and unit standard deviation, or min-max normalization, i.e., scaling the values of each dimension into the range $[0, 1]$. The limitation in z-score or min-max normalization lies in the fact that the statistics of the past observations (during training phase) are used to normalize future observations, which might possess completely different magnitudes due to non-stationarity or concept drift. In order to tackle this problem, several sophisticated methods have been proposed \cite{shao2015self, nayak2014impact}. In addition, hand-crafted stationary features, econometric or quantitative indicators with mathematical assumptions of the underlying processes are also widely used. These financial indicators can sometimes perform relatively well after a long process of experimentation and validation, which, however, prevents their practical implementation in HFT \cite{ntakaris2019feature}.

Different from the aforementioned model-based approaches, data-driven normalization methods aim to directly estimate relevant statistics which are specific to the given analysis task in an end-to-end manner. That is, the normalization step is implemented as a neural network layer whose parameters are jointly optimized with other layers via stochastic gradient descend. Perhaps the most widely used formulation is Batch Normalization (BN) \cite{ioffe2015batch}, which was originally proposed for visual data. BN, however, is mostly used between hidden layers to reduce internal covariate shifts. Proposed for the task of visual style transfer, Instance Normalization (IN) \cite{ulyanov2016instance} was very successful in normalizing the constrast level of generated images. For time-series, an input normalization layer that learns to adaptively estimate the normalization statistics in a given time-series, which outperforms existing schemes, was proposed in \cite{passalis2019deep}.

Existing data-driven approaches, however, neglect the tensor structure inherent in multivariate time-series, performing normalization only along the temporal mode of time-series. In order to take advantage of the tensor representation, the authors in \cite{tran2018temporal} proposed TABL networks which separately capture linear dependency along the temporal and feature dimension in each layer. Since TABL network performs a sequence of weighted sum alternating between the temporal and feature dimension, we propose a data-driven normalization strategy that takes into account statistics from both temporal and spatial dimensions, which is dubbed as Bilinear Normalization (BiN). Combining BiN with TABL, we show that BiN-TABL networks significantly outperforms TABL networks using other normalization strategies in the mid-price movement prediction problem using a large scale Limit Order Book dataset. 

The remainder of the paper is organized as follows. Section 2 reviews related literature in data normalization. In Section 3, we describe the motivation and processing steps of our Bilinear Normalization layer. In Section 4, we provide information about experiment setup, present and discuss our empirical results. Section 5 concludes our work.

\section{Related Work}
Deep neural networks have seen significant improvement over the past decades thanks to the advancement in both hardware and algorithms. On the algorithmic side, training deep networks comprising of multiple layers can be challenging since the distribution of each layer's inputs can change significantly during the iterative optimization process, which harms the error feedback signals. Thus, by manipulating the statistics between layers, we have seen great improvements in optimizing deep neural networks. An early example is the class of initialization methods \cite{glorot2010understanding, he2015delving}, which initialize the network's parameters based on each layer's statistics. However, most of the initialization methods are data independent. A more active approach is the direct manipulation of the statistics by learning them jointly with the network's parameters with the early work called Batch Normalization (BN) \cite{ioffe2015batch}. BN estimates global mean and variance of input data by gradually accumulating the mini-batch statistics. After standardizing the data to have zero-mean and unit variance, BN also learns to scale and shift the distribution. Instead of mini-batch statistics, Instance Normalization \cite{ulyanov2016instance} uses sample-level statistics and learns how to normalize each image so that its contrast matches with that of a predefined style image in the visual style transfer problems. Both BN and IN were originally proposed for visual data, although BN has also been widely used in NLP. 

We are not aware of any data-driven normalization scheme for time-series, except the recently proposed Deep Adaptive Input Normalization (DAIN) formulation \cite{passalis2019deep}, which applies normalization to the input time-series via a 3-stage procedure. Specifically, let $\{\mathbf{X}^{(i)} \in \mathbb{R}^{D \times T}; i=1, \dots, N\}$ be a collection $N$ time-series where $T$ denotes the temporal dimension and $D$ denotes the spatial/feature dimension. In addition, we denote $\mathbf{x}_2^{(i)}(t) \in \mathbb{R}^{D}$ the representation (temporal slice) at time instance $t$ of series $i$. Here the subscript denotes the tensor mode (1 for feature slices and 2 for temporal slices). DAIN first shifts the input time-series by:

\begin{equation}\label{eq1}
\begin{aligned}
& \mathbf{y}_2^{(i)}(t) =  \mathbf{x}_2^{(i)}(t) - \bm{\alpha}^{(i)} \\
& \bm{\alpha}^{(i)} = \mathbf{W}_{a} \mathbf{a}^{(i)} \\
&\mathbf{a}^{(i)} = \frac{1}{T} \sum_{t=1}^{T} \mathbf{x}_2^{(i)}(t) 
\end{aligned}
\end{equation}
where $\mathbf{W}_{a} \in \mathbb{R}^{D \times D}$ is a learnable weight matrix that estimates the amount of shifting from the mean temporal slice ($\mathbf{a}^{(i)}$) calculated from each series. 

After shifting, the intermediate representation $\mathbf{y}_2^{(i)}(t)$ is then scaled as follows:
\begin{equation}\label{eq2}
\begin{aligned}
& \mathbf{z}_2^{(i)}(t) =  \mathbf{y}_2^{(i)}(t) \varoslash \bm{\beta}^{(i)} \\
& \bm{\beta}^{(i)} = \mathbf{W}_{b} \mathbf{b}^{(i)} \\
& \mathbf{b}^{(i)} = \sqrt{\frac{1}{T} \sum_{t=1}^{T} \big(\mathbf{y}_2^{(i)}(t) \odot \mathbf{y}_2^{(i)}(t)\big)}
\end{aligned}
\end{equation}
where $\mathbf{W}_{b} \in \mathbb{R}^{D \times D}$ is a learnable weight matrix that estimates the amount of scaling from the standard deviation ($\mathbf{b}^{(i)}$) along the temporal dimension. In Eq. (\ref{eq2}), the square-root operator is applied element-wise; $\odot$ and $\varoslash$ denote the element-wise multiplication and division, respectively. 

The final step in DAIN is gating, which aims to suppress irrelevant features:
\begin{equation}\label{eq3}
\begin{aligned}
\tilde{\mathbf{x}}_2^{(i)}(t) &=  \mathbf{z}_2^{(i)}(t) \odot \bm{\gamma}^{(i)} \\
\bm{\gamma}^{(i)} & = \mathrm{sigmoid}\big(\mathbf{W}_{c} \mathbf{c}^{(i)} + \mathbf{W}_d \big) \\
\mathbf{c}^{(i)} &= \frac{1}{T} \sum_{t=1}^{T} \mathbf{z}_2^{(i)}(t) 
\end{aligned}
\end{equation}
where $\mathbf{W}_c \in \mathbb{R}^{D\times D}$ and $\mathbf{W}_d \in \mathbb{R}^{D}$ are learnable weights. 

Overall, DAIN takes the input time-series $\mathbf{X}^{(i)}$ and outputs its normalized version $\tilde{\mathbf{X}}^{(i)}$ by manipulating its temporal slices. As we will see in the next Section, our BiN formulation is much simpler (requiring few calculations) and more intuitive compared to DAIN when using with TABL networks. 

\section{Bilinear Normalization (BiN)}
Our proposed BiN layer formulation bears some resemblances to DAIN and IN in that BiN also uses sample-level statistics to manipulate the input distribution. That is, each input sample is normalized based on its statistics only. This is different from BN, which uses global statistics calculated and aggregated from mini-batches. BiN differs from DAIN and IN in that we propose to jointly normalize the input time-series along both temporal and feature dimensions, taking into account the property of bilinear projection in TABL networks. 

The core idea in TABL networks is the separate modeling of linear dependency along the temporal and feature dimension. That is, the interactions between temporal slices and feature slices are captured by bilinear projection:

\begin{equation}\label{eq4}
\mathbf{Y}^{(i)} = \mathbf{W}_1 \mathbf{X}^{(i)} \mathbf{W}_2 
\end{equation}
where $\mathbf{W}_1 \in \mathbb{R}^{D_1 \times D}$ and $\mathbf{W}_2 \in \mathbb{R}^{T \times T_1}$ are the projection parameters, and $\mathbf{Y}^{(i)} \in \mathbb{R}^{D_1 \times T_1}$ is the transformed series. 

In Eq. (\ref{eq4}), $\mathbf{W}_2$ linearly combines $T$ temporal slices $\mathbf{x}_2^{(i)}(t)$ ($t=1, \dots, T$) in $\mathbf{X}^{(i)}$. That is, the function of $\mathbf{W}_2$ is to capture linear patterns in local temporal movement. On the other hand, $\mathbf{W}_1$ linearly combines a set of $D$ feature slices $\mathbf{x}_1^{(i)}(d) \in \mathbb{R}^{T}$ ($d=1, \dots, D$), i.e., row vectors of $\mathbf{X}^{(i)}$, to model local interactions among $D$ different univariate series. 

Due to the above property, it is intuitive to shift and scale not only the distribution of temporal slices $\mathbf{x}_2^{(i)}(t)$ but also that of feature slice $\mathbf{x}_1^{(i)}(d)$. To this end, we propose BiN, which can learn how to jointly manipulate the input data distribution along the temporal and feature dimension.

The normalization along the temporal dimension in BiN is described by the following equations:
\begin{subequations}
	\label{eq5}
	\begin{align}
& \bar{\mathbf{x}}_2^{(i)}  = \frac{1}{T} \sum_{t=1}^{T} \mathbf{x}_2^{(i)}(t) \label{eq5.1}\\
& \bm{\sigma}_2^{(i)} = \sqrt{\frac{1}{T} \sum_{t=1}^{T}\big(\mathbf{x}_2^{(i)}(t)- \bar{\mathbf{x}}_2^{(i)}\big) \odot \big(\mathbf{x}_2^{(i)}(t)- \bar{\mathbf{x}}_2^{(i)}\big)} \label{eq5.2}\\
& \mathbf{Z}_2^{(i)} = \big(\mathbf{X}^{(i)} - \bar{\mathbf{x}}_2^{(i)} \mathbf{1}_T^{\textrm{T}}\big) \varoslash \big(\bm{\sigma}_2^{(i)} \mathbf{1}_T^{\textrm{T}}\big) \label{eq5.3}\\
& \tilde{\mathbf{X}}_2^{(i)}  = \big(\bm{\gamma}_2 \mathbf{1}_T^{\textrm{T}}\big)\odot \mathbf{Z}_2^{(i)} + \bm{\beta}_2 \mathbf{1}_T^{\textrm{T}} \label{eq5.4}
	\end{align}
\end{subequations}
where $\bm{\gamma}_2 \in \mathbb{R}^{D}$ and $\bm{\beta}_2 \in \mathbb{R}^D$ are two learnable weight vectors of BiN. In addition, $ \mathbf{1}_T \in \mathbb{R}^{T}$ is a constant vector having all elements equal to one and $\mathbf{1}_T^{\textrm{T}} \in \mathbb{R}^{1\times T}$ is its transpose. 

In short, given an input series, we first calculate the mean temporal slice $\bar{\mathbf{x}}_2^{(i)} \in \mathbb{R}^{D}$ and its standard deviation $\bm{\sigma}_2^{(i)} \in \mathbb{R}^{D}$ as in Eq. (\ref{eq5.1}, \ref{eq5.2}), which are then used to standardize each temporal slice of the input as in Eq. (\ref{eq5.3}) before applying element-wise scaling (using $\bm{\gamma}_2$) and shifting (using $\bm{\beta}_2$) as in Eq. (\ref{eq5.4}). 

In order to interpret the effects of Eq. (\ref{eq5}), we can view the input series $\mathbf{X}^{(i)}$ as the set $\mathcal{T}^{(i)}$ consisting of $T$ temporal slices, i.e., a set of points in $D$-dimensional space. The process in Eq. (\ref{eq5.3}) moves this set of points around the origin and as well as controlling their spread while keeping their arrangement pattern similarly. If we have two input series $\mathbf{X}^{(i)}$ and $\mathbf{X}^{(j)}$ with $\mathcal{T}^{(i)}$ and $\mathcal{T}^{(j)}$ spread and lie in two completely different areas of this $D$-dimensional space but have the same arrangement pattern, without the alignment performed by Eq. (\ref{eq5.3}), we cannot effectively capture the linear or nonlinear\footnote{Nonlinear patterns can be estimated by several piece-wise linear patterns (by setting the second dimension of $\mathbf{W}_2$ larger than 1, i.e., $T_1> 1$)} arrangement patterns of these points using $\mathbf{W}_2$ in Eq. (\ref{eq4}). Here we should note that although BiN applies additional scaling and shifting as in Eq. (\ref{eq5.4}) after the alignment, the values of $\bm{\gamma}_2$ and $\bm{\beta}_2$ are the same for every input series, thus still keeping their alignments. Since $\bm{\gamma}_2$ and $\bm{\beta}_2$ are optimized together with TABL network's parameters, they enable BiN to manipulate the aligned distributions $\mathcal{T}^{(i)}$ to match with the statistics of other layers. 

While the effect of non-stationarity in the temporal mode are often visible and has been heavily studied, its effects when considered from the feature dimension perspective are less obvious. To see this, let us now view the series $\mathbf{X}^{(i)}$ as the set $\mathcal{D}^{(i)}$ of $D$ points (its $D$ feature slices) in a $T$-dimensional space. Let us also take the previous scenario where two series, $\mathbf{X}^{(i)}$ and $\mathbf{X}^{(j)}$, have $\mathcal{T}^{(i)}$ and $\mathcal{T}^{(j)}$ scattered in different regions of a $D$-dimensional co-ordinate system (viewed under the temporal perspective) before the normalization step in Eq. (\ref{eq5}). When $\mathcal{T}^{(i)}$ and $\mathcal{T}^{(j)}$ are very far away, being viewed from the feature perspective, these two series are also likely to possess $\mathcal{D}^{(i)}$ and $\mathcal{D}^{(j)}$ which are distributed in two different regions of a $T$-dimensional co-ordinate system, although having very similar arrangement. This scenario also prevents $\mathbf{W}_1$ in TABL networks to effectively capture the prominent linear/nonlinear patterns existing in the feature dimension of all input series. Thus, BiN also normalizes the input series along the feature dimension as follows:

\begin{subequations}
	\label{eq6}
	\begin{align}
	& \bar{\mathbf{x}}_1^{(i)}  = \frac{1}{D} \sum_{d=1}^{D} \mathbf{x}_1^{(i)}(d) \label{eq6.1}\\
	& \bm{\sigma}_1^{(i)} = \sqrt{\frac{1}{D} \sum_{d=1}^{D}\big(\mathbf{x}_1^{(i)}(d)- \bar{\mathbf{x}}_1^{(i)}\big) \odot \big(\mathbf{x}_1^{(i)}(d)- \bar{\mathbf{x}}_1^{(i)}\big)} \label{eq6.2}\\
	& \mathbf{Z}_1^{(i)} = \big(\mathbf{X}^{(i)} - \mathbf{1}_D (\bar{\mathbf{x}}_1^{(i)})^{\textrm{T}}\big) \varoslash \big(\mathbf{1}_D (\bm{\sigma}_2^{(i)})^{\textrm{T}} \big) \label{eq6.3}\\
	& \tilde{\mathbf{X}}_1^{(i)}  = \big( \mathbf{1}_D \bm{\gamma}_1^{\textrm{T}} \big)\odot \mathbf{Z}_1^{(i)} + \mathbf{1}_D \bm{\beta}_1^{\textrm{T}} \label{eq6.4}
	\end{align}
\end{subequations}
where $\bm{\gamma}_1 \in \mathbb{R}^{T}$ and $\bm{\beta}_1 \in \mathbb{R}^{T}$ are two learnable weights, and the superscript $(.)^{\textrm{T}}$ denotes the transpose operator.

Overall, BiN takes as input the series $\mathbf{X}^{(i)}$ and outputs $\tilde{\mathbf{X}}^{(i)}$, which is the linear combination of $\tilde{\mathbf{X}}_1^{(i)}$ and $\tilde{\mathbf{X}}_2^{(i)}$ from Eq. (\ref{eq6.4}) and (\ref{eq5.4}), respectively:
\begin{equation}\label{eq7}
\tilde{\mathbf{X}}^{(i)} = \lambda_1 \tilde{\mathbf{X}}_1^{(i)} + \lambda_2 \tilde{\mathbf{X}}_2^{(i)}
\end{equation} 

where $\lambda_1 \in \mathbb{R}$ and $\lambda_2 \in \mathbb{R}$ are two learnable scalars, which enables BiN to weigh the importance of temporal and feature normalization. Here we should note that $\lambda_1$ and $\lambda_2$ are constrained to be non-negative. This constraint is achieved during stochastic optimization by setting the value (of $\lambda_1$ or $\lambda_2$) to $0$ whenever the updated value is negative.

\section{Experiments}

\subsection{FI-2010 Limit Order Book Dataset}
In finance, a limit order placed with a bank or a brokerage is a type of trade order to buy or sell a fixed amount of assets with a specified price. In a limit order, the trader must specify three pieces of information: the type (buy or sell), the price of a unit of the asset, and the volume (the number of stock items he or she wants to trade). Basically, with a limit order, the trader only wants to buy or sell an asset at a price that is at least as good as what he or she specifies. That is, with a buy (sell) limit order, the trader only wants to buy (sell) an asset with the price equal or lower (higher) than the price he or she specifies. The buy (bid) and sell (ask) limit orders form two sides of the LOB: the bid and the ask side. At time $t$, the best bid price ($p_b^1(t)$) and the best ask price ($p_a^1(t)$) are defined as the highest bid and the lowest ask price that exist in the LOB, respectively. When a new limit order arrives, the system aggregates and sorts the orders on both sides so that the best bid and best ask orders are placed on the top, which is called the first level. If there are limit orders where the bid price is equal or higher than the lowest ask, i.e. $p_b^1(t) \geq p_a^1(t)$, those orders are immediately fulfilled and removed from the LOB. 

In order to evaluate the proposed BiN layer in the problem of financial forecasting, we conducted empirical analysis on FI-2010 \cite{ntakaris2018benchmark}, a large scale, publicly available Limit Order Book (LOB) dataset, which contains buy and sell limit order information (the prices and volumes) over $10$ business days from $5$ Finnish stocks traded in Helsinki Exchange (operated by NASDAQ Nordic). At each time instance, the dataset contains the prices and volumes from the top $10$ levels of both buy and sell sides, leading to a $40$-dimensional vector representation. 

Using this dataset, we investigated the problem of mid-price movement prediction in the next $H=\{10, 20, 50\}$ events. Mid-price at a given time instance is the mean value between the best bid and best ask prices, which is a virtual quantity since no trade can take place at this price at the given time. Its movements (stationary, increasing, decreasing), however, reflects the dynamic of the LOB and the market. Therefore, being able to predict the future movements of the mid-price using the current and past order information is of great importance. For more information on FI-2010 dataset and the limit order book, we refer the reader to \cite{ntakaris2018benchmark}.

\begin{figure}[t!]
	\centering
	\includegraphics[width=0.8\linewidth]{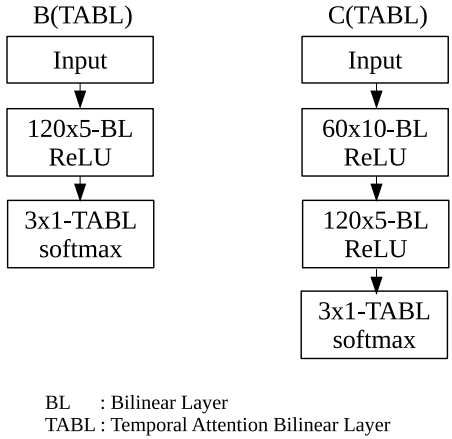}
	\caption{Network Topologies}\label{f1}
\end{figure}

\subsection{Experimental protocols}

We followed the same experimental setup proposed in \cite{tran2018temporal} which used LOB information from the first $7$ days to train the models and the last $3$ days for evaluating purpose. The input to all the models consists of the prices and volumes of the top $10$ levels over the $10$ most recent events, i.e., each input sample is a matrix of size $40\times 10$. The target is the mid-price movement after $H=\{10, 20, 50\}$ events. $H$ is also referred to as the prediction horizon, and here we should note that different models are trained for each value of the prediction horizon.  

Furthermore, we also followed \cite{tran2018temporal} and used the the same TABL architectures that produced the best performances in \cite{tran2018temporal} to evaluate, denoted as B(TABL) and C(TABL) as in \cite{tran2018temporal}. B(TABL) is an architecture that has only one hidden layer while C(TABL) has two hidden layers. The topologies of B(TABL) and C(TABL) are illustrated in Figure \ref{f1}. The results for C(TABL) networks applying our BiN layer and BN layer as an input normalization layer are denoted as BiN-C(TABL) and BN-C(TABL), respectively. 

For weight regularization, we experimented with two types of weight regularization: weight decay with a coefficient of $1\textrm{e}{-3}$ and max-norm constraint with the maximum norm set to $10.0$. After each hidden layer, we also applied dropout regularization with the dropout rate set to $0.1$. ADAM optimizer was used to optimize the networks' parameters. Each model was trained for a total of $80$ epochs with the learning rate starting at $1\textrm{e}{-3}$ and dropping to $1\textrm{e}{-4}$, then to $1\mathrm{e}{-5}$ at epoch $11$ and $71$, respectively. Since the objective is to train each model to predict the future movement of the mid-price, cross-entropy was used as the loss function. Similar to \cite{tran2018temporal, tran2019data}, a weighted cross-entropy loss function was used to counter the effect of data imbalance in the FI-2010 dataset. That is, the loss term associated with each class is multiplied with a constant that is inversely proportional with the number of samples in that class. 

Accuracy, averaged Precision, Recall and F1 are reported as the performance metrics. Since FI-2010 is an imbalanced dataset, we focus our analysis on the F1 measure. In addition, each experiment was run $5$ times and the median value is reported. 

\subsection{Experiment Results}

\begin{table}[t!]
	\begin{center}
		\caption{Experiment Results. Bold-face numbers denote the best F1 measure among the normalization strategies}\label{t1}
		\resizebox{\linewidth}{!}{
			\begin{tabular}{|c|c|c|c|c|}
				\multicolumn{5}{c}{} \\ \hline
				\textbf{Models}		& \textbf{Accuracy \%} 	& \textbf{Precision \%} & \textbf{Recall \%}	& \textbf{F1 \%} 		\\ \hline \hline
				\multicolumn{5}{|c|}{\textit{Prediction Horizon $H=10$}} \\ \hline
				CNN\cite{tsantekidis2017forecasting}		& -			& $50.98$	&$65.54$	& $55.21$	\\ \hline
				LSTM\cite{tsantekidis2017using}	& -			& $60.77$	&$75.92$	& $66.33$	\\ \hline \hline
				C(BL) \cite{tran2018temporal}	& $82.52$	& $73.89$	&$76.22$	& $75.01$ \\ \hline
				DeepLOB \cite{zhang2019deeplob}	& $84.47$	& $84.00$	&$84.47$	& $83.40$ \\ \hline \hline
				DAIN-MLP \cite{passalis2019deep}	& -	& $65.67$	&$71.58$	& $68.26$ \\ \hline
				DAIN-RNN \cite{passalis2019deep}	& -	& $61.80$	&$70.92$	& $65.13$ \\ \hline
				C(TABL)	\cite{tran2018temporal} & $84.70$	& $76.95$	&$78.44$	& $77.63$ \\ \hline 
				BN-C(TABL)  & $79.20$	& $68.48$	&$72.36$	& $66.87$ \\ \hline
				BiN-C(TABL)	& $86.87$	& $80.29$	&$81.84$	& $\mathbf{81.04}$ \\ \hline \hline

				\multicolumn{5}{|c|}{\textit{Prediction Horizon $H=20$}} \\ \hline
				CNN\cite{tsantekidis2017forecasting}		& -			& $54.79$	&$67.38$	& $59.17$	\\ \hline
				LSTM\cite{tsantekidis2017using}	& -			& $59.60$	&$70.52$	& $62.37$	\\ \hline 
				C(BL) \cite{tran2018temporal}	& $72.05$	& $65.04$	&$65.23$	& $64.89$ \\ \hline 
				DeepLOB	\cite{zhang2019deeplob} & $74.85$	& $74.06$	&$74.85$	& $72.82$ \\ \hline \hline
				DAIN-MLP \cite{passalis2019deep}	& -	& $62.10$	&$70.48$	& $65.31$ \\ \hline
				DAIN-RNN \cite{passalis2019deep}	& -	& $59.16$	&$68.51$	& $62.03$ \\ \hline
				C(TABL)	\cite{tran2018temporal} & $73.74$	& $67.18$	&$66.94$	& $66.93$ \\ \hline
				BN-C(TABL)  & $70.70$	& $63.10$	&$63.78$	& $63.43$ \\ \hline
				BiN-C(TABL)	& $77.28$	& $72.12$	&$70.44$	& $\mathbf{71.22}$ \\ \hline \hline
				
				\multicolumn{5}{|c|}{\textit{Prediction Horizon $H=50$}} \\ \hline
				CNN\cite{tsantekidis2017forecasting}		& -			& $55.58$	&$67.12$	& $59.44$	\\ \hline
				LSTM\cite{tsantekidis2017using}	& -			& $60.03$	&$68.58$	& $61.43$	\\ \hline 
				C(BL) \cite{tran2018temporal} 	& $78.96$	& $77.85$	&$77.04$	& $77.40$ \\ \hline
				DeepLOB \cite{zhang2019deeplob}	& $80.51$	& $80.38$	&$80.51$	& $80.35$ \\ \hline \hline
				DAIN-MLP \cite{passalis2019deep}	& -	& -	& -	& - \\ \hline
				DAIN-RNN \cite{passalis2019deep}	& -	& -	& -	& -  \\ \hline
				C(TABL) \cite{tran2018temporal}	& $79.87$	& $79.05$	&$77.04$	& $78.44$ \\ \hline
				BN-C(TABL)  	& $77.16$	& $75.70$	&$75.04$	& $75.34$ \\ \hline
				BiN-C(TABL)	& $88.54$	& $89.50$	&$86.99$	& $\mathbf{88.06}$ \\ \hline
			\end{tabular}
		}
	\end{center}
\end{table}

Table \ref{t1} shows the experiment results in three prediction horizons $H=\{10, 20, 50\}$ of the proposed BiN-C(TABL) networks in comparisons with the original TABL architecture C(TABL), other input normalization strategies BN-C(TABL), DAIN-MLP, DAIN-RNN (the lower section of each horizon) as well as recent state-of-the-art results for deep architectures (the upper section). 

\begin{table}[t!]
	\begin{center}
		\caption{Improvement comparison between BiN-C(TABL) versus BiN-B(TABL)}\label{t3}
		\resizebox{\linewidth}{!}{
			\begin{tabular}{|c|c|c|c|c|}
				\multicolumn{5}{c}{} \\ \hline
				\textbf{Models}		& \textbf{Accuracy \%} 	& \textbf{Precision \%} & \textbf{Recall \%}	& \textbf{F1 \%} 		\\ \hline \hline
				\multicolumn{5}{|c|}{\textit{Prediction Horizon $H=10$}} \\ \hline
				B(TABL)	\cite{tran2018temporal}	& $78.91$	& $68.04$	&$71.21$	& $69.20$ \\ \hline
				C(TABL)	\cite{tran2018temporal} & $84.70$	& $76.95$	&$78.44$	& $77.63$ \\ \hline \hline
				BiN-B(TABL) & $86.92$ 	& $80.43$	&$81.82$	& $81.10$ \\ \hline
				BiN-C(TABL)	& $86.87$	& $80.29$	&$81.84$	& $81.04$ \\ \hline \hline
				\multicolumn{5}{|c|}{\textit{Prediction Horizon $H=20$}} \\ \hline
				B(TABL) \cite{tran2018temporal}       & $70.80$       & $63.14$       &$62.25$        & $62.22$ \\ \hline \hline
				C(TABL)	\cite{tran2018temporal} & $73.74$	& $67.18$	&$66.94$	& $66.93$ \\ \hline \hline
				BiN-B(TABL) & $77.54$	& $72.56$	&$70.22$	& $71.29$ \\ \hline
				BiN-C(TABL)	& $77.28$	& $72.12$	&$70.44$	& $71.22$ \\ \hline \hline 
				\multicolumn{5}{|c|}{\textit{Prediction Horizon $H=50$}} \\ \hline
				B(TABL) \cite{tran2018temporal} & $75.58$       & $74.58$       &$73.09$        & $73.64$ \\ \hline
				C(TABL) \cite{tran2018temporal}	& $79.87$	& $79.05$	&$77.04$	& $78.44$ \\ \hline \hline
				BiN-B(TABL) & $88.44$	& $89.36$	&$86.92$	& $87.96$ \\ \hline
				BiN-C(TABL)	& $88.54$	& $89.50$	&$86.99$	& $88.06$ \\ \hline
			\end{tabular}
		}
	\end{center}
\end{table}

It is clear that our proposed BiN layer (BiN-C(TABL)) when used to normalize the input data yields significant improvement over the original TABL networks (C(TABL)) in all prediction horizons. Especially, for the longest horizon $H=50$, BiN enhances the C(TABL) network with up to $10\%$ improvement (from $78.44\%$ to $88.06\%$) in average F1 measure. Compared with DAIN, the performances achieved by our normalization strategy coupled with TABL networks far exceed that of DAIN coupled with MLP or RNN. Regarding BN when used as an input normalization scheme, it is obvious that BN deteriorates the performance of the C(TABL) network. For example, in case of $H=10$, adding BN to C(TABL) network leads to more than $10\%$ drop in averaged F1. This behaviour is expected since BN was originally designed to reduce covariate shift between hidden layers of Convolutional Neural Network, rather than as a mechanism to normalize multivariate time-series. 

Comparing BiN-C(TABL) with the state-of-the-arts CNN-LSTM architecture having 11 hidden layers called DeepLOB \cite{zhang2019deeplob}, it is clear that our proposed normalization layer helps a TABL network having only 2 hidden layers to significantly close the gaps when $H=10$ and $H=20$ ($81.04\%$ versus $83.40\%$ for $H=10$, and $71.22\%$ versus $72.82\%$ for $H=20$), while outperforming DeepLOB by a large margin when $H=50$ ($88.06\%$ versus $80.35\%$). 

In order to investigate the extent of improvement with respect to different TABL network topologies when using our proposed normalization layer, we also conducted experiments with a smaller TABL network, namely B(TABL) as proposed in \cite{tran2018temporal}. B(TABL) has only one hidden layer with a total of $5843$ parameters, compared to C(TABL) which has two hidden layers with a total of $11343$ parameters. The results are shown in Table \ref{t3}. First of all, it is obvious that adding the proposed normalization layer significantly enhances both B(TABL) and C(TABL) in different prediction horizons. More surprisingly, BiN-B(TABL) networks perform as well as BiN-C(TABL) networks in all prediction horizons, making the additional parameters in BiN-C(TABL) redundant. Here we should note that adding our proposed BiN normalization layer to B(TABL) networks only leads to a mere increase of $102$ parameters while achieving the same performances as BiN-C(TABL) networks, which have double the amount of parameters. 

Since BN has been widely used for hidden layers, we also compared the performance of BiN and BN when applied to all layers in Table \ref{t2}. The upper section of each horizon shows the performance of BiN and BN when applied only to the input layer while the lower section shows their performance when applied to all layers. As we can see from Table \ref{t2}, there is virtually no differences between the two arrangements. This result shows that adding normalization to the hidden layers bring no improvement to both strategies and the improvements obtained for TABL networks are indeed attributed to the input data normalization performed by BiN.

\begin{table}[t!]
	\begin{center}
		\caption{Comparisons between Bilinear Normalization and Batch Normalization when applied to only input layer (BiN-C(TABL) and BN-C(TABL)) or all layers (BiN-C(TABL)-BiN and BN-C(TABL)-BN}\label{t2}
		\resizebox{\linewidth}{!}{
			\begin{tabular}{|c|c|c|c|c|}
				\multicolumn{5}{c}{} \\ \hline
				\textbf{Models}		& \textbf{Accuracy \%} 	& \textbf{Precision \%} & \textbf{Recall \%}	& \textbf{F1 \%} 		\\ \hline \hline
				\multicolumn{5}{|c|}{\textit{Prediction Horizon $H=10$}} \\ \hline
				BN-C(TABL)  & $79.20$	& $68.48$	&$72.36$	& $66.87$ \\ \hline
				BiN-C(TABL)	& $86.87$	& $80.29$	&$81.84$	& $81.04$ \\ \hline \hline
				BN-C(TABL)-BN  & $78.72$	& $68.02$	&$72.58$	& $69.98$ \\ \hline
				BiN-C(TABL)-BiN & $86.84$	& $80.25$	&$81.85$	& $81.03$ \\ \hline \hline

				\multicolumn{5}{|c|}{\textit{Prediction Horizon $H=20$}} \\ \hline
				BN-C(TABL)  & $70.70$	& $63.10$	&$63.78$	& $63.43$ \\ \hline
				BiN-C(TABL)	& $77.28$	& $72.12$	&$70.44$	& $71.22$ \\ \hline \hline
				BN-C(TABL)-BN  & $71.28$	& $63.77$	&$63.65$	& $63.75$ \\ \hline
				BiN-C(TABL)-BiN & $76.68$	& $71.15$	&$70.48$	& $70.80$ \\ \hline \hline
				\multicolumn{5}{|c|}{\textit{Prediction Horizon $H=50$}} \\ \hline
				BN-C(TABL)  	& $77.16$	& $75.70$	&$75.04$	& $75.34$ \\ \hline
				BiN-C(TABL)	& $88.54$	& $89.50$	&$86.99$	& $88.06$ \\ \hline \hline
				BN-C(TABL)-BN  	& $76.74$	& $75.34$	&$74.66$	& $74.97$ \\ \hline
				BiN-C(TABL)-BiN	& $88.44$	& $89.36$	&$86.92$	& $87.96$ \\ \hline
			\end{tabular}
		}
	\end{center}
\end{table}

\section{Conclusions}
In this paper, we propose BiN, a data-driven time-series normalization strategy which is designed to tackle the potential difficulties posed by noisy, non-stationary financial time-series. Our proposed normalization layer takes into account the property of bilinear projection in TABL networks and aligns the multivariate time-series in both feature and temporal dimensions. Using a large scale limit order book dataset focusing on stock movement prediction, we demonstrated that BiN can greatly enhances the performances of previous state-of-the-arts TABL networks while requiring few additional parameters and computation.

\bibliography{reference}

\begin{thebibliography}{10}

\bibitem{tran2018temporal}
D.~T. Tran, A.~Iosifidis, J.~Kanniainen, and M.~Gabbouj, ``Temporal
  attention-augmented bilinear network for financial time-series data
  analysis,'' {\em IEEE transactions on neural networks and learning systems},
  vol.~30, no.~5, pp.~1407--1418, 2018.

\bibitem{zhang2019deeplob}
Z.~Zhang, S.~Zohren, and S.~Roberts, ``Deeplob: Deep convolutional neural
  networks for limit order books,'' {\em IEEE Transactions on Signal
  Processing}, vol.~67, no.~11, pp.~3001--3012, 2019.

\bibitem{passalis2018temporal}
N.~Passalis, A.~Tefas, J.~Kanniainen, M.~Gabbouj, and A.~Iosifidis, ``Temporal
  bag-of-features learning for predicting mid price movements using high
  frequency limit order book data,'' {\em IEEE Transactions on Emerging Topics
  in Computational Intelligence}, 2018.

\bibitem{tran2017multilinear}
D.~T. Tran, M.~Gabbouj, and A.~Iosifidis, ``Multilinear class-specific
  discriminant analysis,'' {\em Pattern Recognition Letters}, 2017.

\bibitem{tran2017tensor}
D.~T. Tran, M.~Magris, J.~Kanniainen, M.~Gabbouj, and A.~Iosifidis, ``Tensor
  representation in high-frequency financial data for price change
  prediction,'' {\em IEEE Symposium Series on Computational Intelligence
  (SSCI)}, 2017.

\bibitem{passalis2017time}
N.~Passalis, A.~Tsantekidis, A.~Tefas, J.~Kanniainen, M.~Gabbouj, and
  A.~Iosifidis, ``Time-series classification using neural bag-of-features,'' in
  {\em Signal Processing Conference (EUSIPCO), 2017 25th European},
  pp.~301--305, IEEE, 2017.

\bibitem{ren2015faster}
S.~Ren, K.~He, R.~Girshick, and J.~Sun, ``Faster r-cnn: Towards real-time
  object detection with region proposal networks,'' in {\em Advances in neural
  information processing systems}, pp.~91--99, 2015.

\bibitem{redmon2016you}
J.~Redmon, S.~Divvala, R.~Girshick, and A.~Farhadi, ``You only look once:
  Unified, real-time object detection,'' in {\em Proceedings of the IEEE
  conference on computer vision and pattern recognition}, pp.~779--788, 2016.

\bibitem{iosifidis2012view}
A.~Iosifidis, A.~Tefas, and I.~Pitas, ``View-invariant action recognition based
  on artificial neural networks,'' {\em IEEE transactions on neural networks
  and learning systems}, vol.~23, no.~3, pp.~412--424, 2012.

\bibitem{tran2018improving}
D.~T. Tran, A.~Iosifidis, and M.~Gabbouj, ``Improving efficiency in
  convolutional neural networks with multilinear filters,'' {\em Neural
  Networks}, vol.~105, pp.~328--339, 2018.

\bibitem{tran2019multilinear}
D.~T. Tran, M.~Yamac, A.~Degerli, M.~Gabbouj, and A.~Iosifidis, ``Multilinear
  compressive learning,'' {\em arXiv preprint arXiv:1905.07481}, 2019.

\bibitem{tran2019bmultilinear}
D.~T. Tran, M.~Gabbouj, and A.~Iosifidis, ``Multilinear compressive learning
  with prior knowledge,'' {\em arXiv preprint arXiv:2002.07203}, 2020.

\bibitem{bahdanau2014neural}
D.~Bahdanau, K.~Cho, and Y.~Bengio, ``Neural machine translation by jointly
  learning to align and translate,'' {\em arXiv preprint arXiv:1409.0473},
  2014.

\bibitem{devlin2018bert}
J.~Devlin, M.-W. Chang, K.~Lee, and K.~Toutanova, ``Bert: Pre-training of deep
  bidirectional transformers for language understanding,'' {\em arXiv preprint
  arXiv:1810.04805}, 2018.

\bibitem{shao2015self}
X.~Shao, ``Self-normalization for time series: a review of recent
  developments,'' {\em Journal of the American Statistical Association},
  vol.~110, no.~512, pp.~1797--1817, 2015.

\bibitem{nayak2014impact}
S.~Nayak, B.~B. Misra, and H.~S. Behera, ``Impact of data normalization on
  stock index forecasting,'' {\em Int. J. Comp. Inf. Syst. Ind. Manag. Appl},
  vol.~6, pp.~357--369, 2014.

\bibitem{ntakaris2019feature}
A.~Ntakaris, G.~Mirone, J.~Kanniainen, M.~Gabbouj, and A.~Iosifidis, ``Feature
  engineering for mid-price prediction with deep learning,'' {\em Ieee Access},
  vol.~7, pp.~82390--82412, 2019.

\bibitem{ioffe2015batch}
S.~Ioffe and C.~Szegedy, ``Batch normalization: Accelerating deep network
  training by reducing internal covariate shift,'' {\em arXiv preprint
  arXiv:1502.03167}, 2015.

\bibitem{ulyanov2016instance}
D.~Ulyanov, A.~Vedaldi, and V.~Lempitsky, ``Instance normalization: The missing
  ingredient for fast stylization,'' {\em arXiv preprint arXiv:1607.08022},
  2016.

\bibitem{passalis2019deep}
N.~Passalis, A.~Tefas, J.~Kanniainen, M.~Gabbouj, and A.~Iosifidis, ``Deep
  adaptive input normalization for price forecasting using limit order book
  data,'' {\em arXiv preprint arXiv:1902.07892}, 2019.

\bibitem{glorot2010understanding}
X.~Glorot and Y.~Bengio, ``Understanding the difficulty of training deep
  feedforward neural networks,'' in {\em Proceedings of the thirteenth
  international conference on artificial intelligence and statistics},
  pp.~249--256, 2010.

\bibitem{he2015delving}
K.~He, X.~Zhang, S.~Ren, and J.~Sun, ``Delving deep into rectifiers: Surpassing
  human-level performance on imagenet classification,'' in {\em Proceedings of
  the IEEE international conference on computer vision}, pp.~1026--1034, 2015.

\bibitem{ntakaris2018benchmark}
A.~Ntakaris, M.~Magris, J.~Kanniainen, M.~Gabbouj, and A.~Iosifidis,
  ``Benchmark dataset for mid-price forecasting of limit order book data with
  machine learning methods,'' {\em Journal of Forecasting}, vol.~37, no.~8,
  pp.~852--866, 2018.

\bibitem{tran2019data}
D.~T. Tran, J.~Kanniainen, M.~Gabbouj, and A.~Iosifidis, ``Data-driven neural
  architecture learning for financial time-series forecasting,'' {\em arXiv
  preprint arXiv:1903.06751}, 2019.

\bibitem{tsantekidis2017forecasting}
A.~Tsantekidis, N.~Passalis, A.~Tefas, J.~Kanniainen, M.~Gabbouj, and
  A.~Iosifidis, ``Forecasting stock prices from the limit order book using
  convolutional neural networks,'' in {\em Business Informatics (CBI), 2017
  IEEE 19th Conference on}, vol.~1, pp.~7--12, IEEE, 2017.

\bibitem{tsantekidis2017using}
A.~Tsantekidis, N.~Passalis, A.~Tefas, J.~Kanniainen, M.~Gabbouj, and
  A.~Iosifidis, ``Using deep learning to detect price change indications in
  financial markets,'' in {\em Signal Processing Conference (EUSIPCO), 2017
  25th European}, pp.~2511--2515, IEEE, 2017.

\end{thebibliography}
\bibliographystyle{ieeetr}

\end{document}